\documentclass[preprint,12pt]{emulateapj}
\pdfoutput=1  

\usepackage{apjfonts}
\usepackage{color}
\usepackage{natbib}
\usepackage{amssymb,amsmath}
\usepackage[pdftex,           
bookmarks=true,                   
bookmarksnumbered=true,  
colorlinks=true,      
citecolor=blue,                     
linkcolor=blue,              
menucolor=blue,                
urlcolor=cyan,                       
frenchlinks=False]{hyperref}

\providecommand{\eprint}[1]{\href{http://arxiv.org/abs/#1}{#1}}
\providecommand{\adsurl}[1]{\href{#1}{ADS}}

\shortauthors{Albrecht et al.\ 2011} \shorttitle{The rotation of the
  oblique star WASP-7}
\begin{document}

\title{A High Stellar Obliquity in the WASP-7 Exoplanetary System}

\author{
  Simon Albrecht\altaffilmark{1}, 
  Joshua N.\ Winn\altaffilmark{1},
  R.\ Paul Butler\altaffilmark{2},
  Jeffrey D.\ Crane\altaffilmark{3},
  Stephen A.\ Shectman\altaffilmark{3},
  Ian B.\ Thompson\altaffilmark{3},
  Teruyuki Hirano\altaffilmark{1,4},
  Robert A.\ Wittenmyer\altaffilmark{5}
}

\altaffiltext{1}{Department of Physics, and Kavli Institute for
  Astrophysics and Space Research, Massachusetts Institute of
  Technology, Cambridge, MA 02139, USA}

\altaffiltext{2}{Department of Terrestrial Magnetism, Carnegie
  Institution of Washington, 5241 Broad Branch Road NW, Washington, DC
  20015, USA}
 
\altaffiltext{3}{The Observatories of the Carnegie Institution of
  Washington, 813 Santa Barbara Street, Pasadena, CA 91101, USA}

\altaffiltext{4}{Department of Physics, The University of Tokyo, Tokyo 113-0033, Japan}
 
\altaffiltext{5}{Department of Astrophysics, School of Physics,
  University of NSW, 2052, Australia}

\altaffiltext{$\star$}{The data presented herein were collected with
  the the Magellan (Clay) Telescope located at Las Campanas
  Observatory, Chile.}

\begin{abstract}

  We measure a tilt of $86\pm6^{\circ}$ between the sky projections of
  the rotation axis of the WASP-7 star, and the orbital axis of its
  close-in giant planet. This measurement is based on observations of
  the Rossiter-McLaughlin (RM) effect with the Planet Finder
  Spectrograph on the Magellan~II telescope. The result conforms with
  the previously noted pattern among hot-Jupiter hosts, namely, that
  the hosts lacking thick convective envelopes have high obliquities.
  Because the planet's trajectory crosses a wide range of stellar
  latitudes, observations of the RM effect can in principle reveal the
  stellar differential rotation profile; however, with the present
  data the signal of differential rotation could not be detected. The
  host star is found to exhibit radial-velocity noise (``stellar
  jitter'') with an amplitude of $\approx$30~m~s$^{-1}$ over a
  timescale of days.\\\\

\end{abstract}


\section{Introduction}
\label{sec:intro}

In the solar system the Sun's equatorial plane is aligned to within
$7^{\circ}$ with the ecliptic. For many stars in systems with close-in
gas giants (``hot-Jupiters''), this is not the case. Over the last
three years it was found that some hot Jupiters have highly inclined
or even retrograde orbits with respect to the rotational spins of
their host stars \cite[see
e.g.][]{hebrard2008,winn2009,johnson2009,triaud2010,winn2010b,simpson2011}.
Understanding what causes these orbital tilts and why some hot
Jupiters are well-aligned with their parent stars might aid our
understanding of why these giant planets are found so close to their host
stars, compared to Jupiter.

Different classes of processes have been proposed which might
transport giant planets from their presumed birthplaces at distances
of many astronomical units from their host stars, inward to a fraction
of an astronomical unit, where we find them. Some of these processes
are expected to change the relative orientation between the stellar
and orbital spin
\cite[e.g.][]{nagasawa2008,fabrycky2007,chatterjee2008}, while others
will conserve the relative orientation between orbital and stellar
spin \cite[e.g.][]{lin1996}, or even reduce a misalignment between
them \citep{cresswell2007}.

\cite{winn2010} and \cite{schlaufman2010} found that close in giant
planets tend to have orbits aligned with the stellar spin if the
effective temperature ($T_{\rm eff}$) of their host star is $\lesssim
6250$~K and misaligned otherwise. \cite{winn2010} further speculated
that this might indicate that {\it all} giant planets are transported
inward by processes that create large obliquities. In this picture,
tidal torques exerted on the star by the close in planet realign the
two angular momentum vectors. The realignment time scale would be
short for planets around stars with convective envelopes ($T_{\rm eff}
\lesssim 6250$K), but long if the star does not have a convective
envelope ($\gtrsim 6250$K). Adding to this picture, \cite{triaud2011}
recently argued that relatively young stars have high obliquities,
while older stars are observed to have low obliquities.

Here we present measurements of the spin-orbit angle in the WASP-7
system. The planet WASP-7b was discovered by \cite{hellier2009} and
found to have a mass of 0.96\,M$_{\rm Jup}$. The host star has a mass
of 1.28\,$M_{\odot}$, a projected rotation speed of
$17\pm2$~km\,s$^{-1}$ \citep{hellier2009}, an effective temperature of
$6520 \pm 70$\,K, and a solar metallicity ([Fe/H]~$= 0.0\pm 0.1$)
\citep{maxted2011}. Based on the aforementioned pattern, we would
expect that our measurement would show a misalignment between orbital
and stellar spins.

This article is organized as follows. The following section describes
the new spectroscopic data, and the analysis of the
Rossiter-McLaughlin effect. Section \ref{sec:jitter_ecc} considers
some possible explanations for the high level of noise in the radial
velocities, including the possibility of an eccentric orbit.  Section
\ref{sec:discussion} discusses the impact of the radial-velocity noise
on the measurement of the stellar obliquity.  This section also
presents an attempt to detect the differential rotation of the host
star using the Rossiter-McLaughlin effect.

\section{Observations}
\label{sec:obs}

We observed WASP-7 with the Magellan~II (Clay) 6.5\,m telescope and
the Planet Finder Spectrograph (PFS; \citealt{crane2010}).  We
gathered 37 spectra spanning the transit of 2010~August~27/28. The
integration times were $10$\,min and the complete sequence spanned
$\sim7.5$~hr. The stellar spectra were observed through an iodine gas
cell, imprinting a dense forest of sharp absorption lines on the
stellar spectra to help establish the wavelength scale and
instrumental profile. During the transit night, an additional spectrum
was obtained without the iodine cell, to serve as a template spectrum
for relative radial-velocity (RV) determination.

\section{Analysis of the Rossiter-McLaughlin effect}
\label{sec:analysis}

\begin{figure} 
  \begin{center} 
    \includegraphics[width=9.5cm]{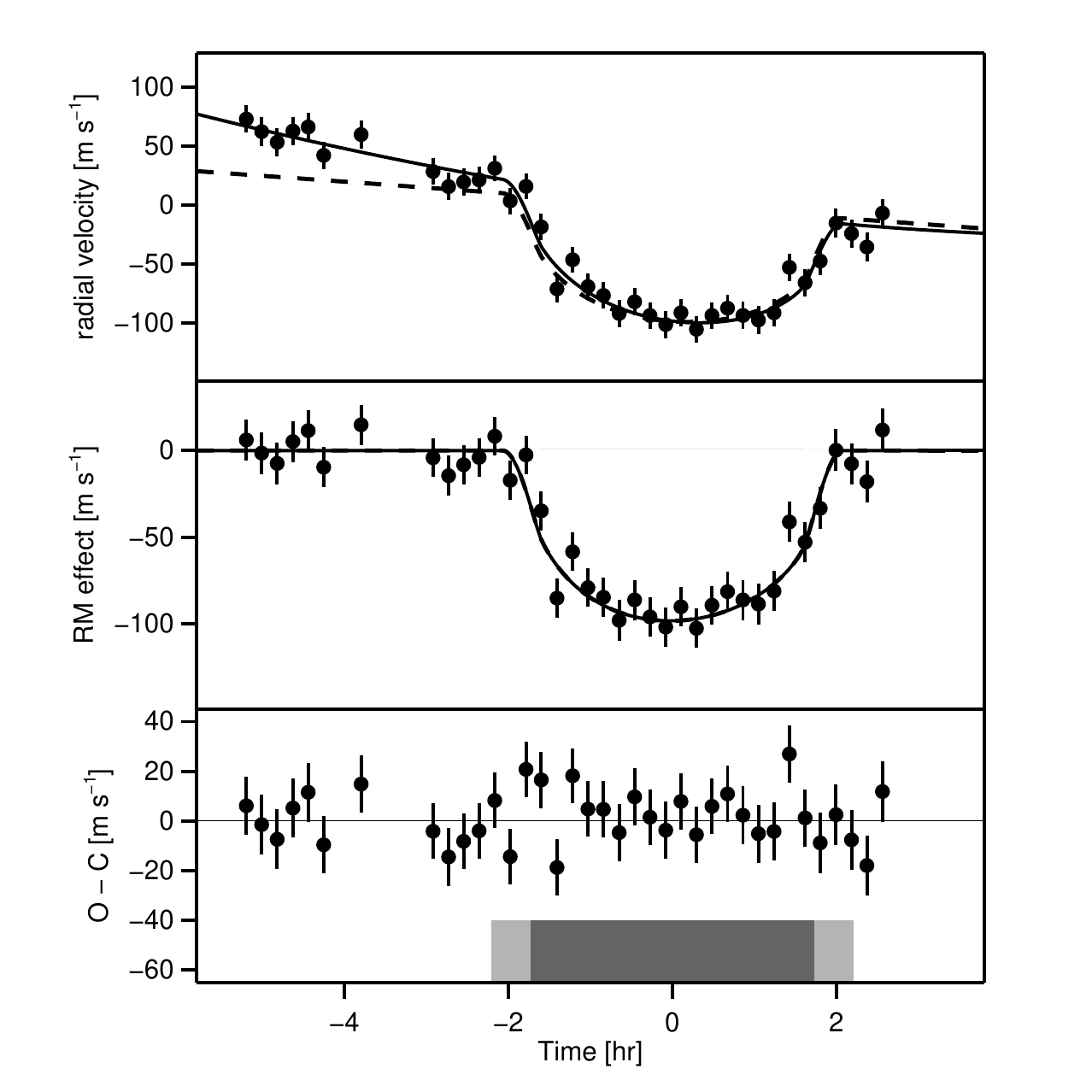}
    \caption {\label{fig:wasp7_rv} {\bf Radial velocities of WASP-7
        before, during, and after the transit of its planet.}  The
      radial velocities are plotted as a function of time from
      inferior conjunction. The upper panel shows the measured RVs
      and the best-fitting model. The dashed line shows the same RM
      model, but with an orbital model with parameters fixed at those
      presented by \cite{hellier2009} (see also
      section~\ref{sec:jitter_ecc}). In the middle panel, the apparent
      orbital contribution to the observed RVs has been subtracted,
      thereby isolating the RM effect. The lower panel shows the
      residuals. The light and dark gray bars in the lowest panel
      indicate times of first, second, third, and fourth contact.}
  \end{center}
\end{figure}

To derive the relative RVs we compared the spectra observed through
the iodine cell with the stellar template spectrum multiplied by an
iodine template spectrum. The velocity shift of the stellar template
as well as the parameters of the point-spread function (PSF) of the
spectrograph are free parameters in this comparison. The velocity
shift of the template that gives the best fit to an observed spectrum
represents the measured relative RV. In particular we used a code
based on that of \cite{butler1996}. The RVs are presented in
Table~\ref{tab:wasp7_rv} and displayed in Figure~\ref{fig:wasp7_rv}.

We take advantage of the Rossiter-McLaughlin (RM) effect to measure
the projected angle between the orbital and stellar spins ($\lambda$)
and the projected stellar rotation speed ($v \sin i_{\star}$). Here
$v$ indicates the stellar rotation speed and $i_{\star}$ the
inclination of the stellar spin axis towards the observer. The RM
effect is a spectroscopic distortion of the rotationally-broadened
stellar absorption lines, which occurs when a companion star or planet
is in front of the star and hides part of the rotating stellar surface
from the observer's view. The position of the distortion on the
stellar absorption line depends on the radial velocity of the hidden
portion of the stellar surface. Therefore the observed distortion of
the stellar absorption lines can be connected to the geometry of the
transit. This shape change can be measured directly and relevant
parameters can be derived \citep{albrecht2007,cameron2010}.

Our analysis is divided into 7 parts. In section~\ref{sec:qualitative}
we discuss the data qualitatively.
Sections~\ref{sec:broaden}--\ref{sec:rv_var} discuss different
phenomena which affect the shape of stellar absorption lines.
Section~\ref{sec:quantitative} discusses the model of the RM effect
which we adopted, and presents the quantitative results.

\begin{table}
  \caption{Relative Radial Velocity measurements of WASP-7}
  \label{tab:wasp7_rv}
  \begin{center}
    \smallskip
    \begin{tabular}{l c c }
      \hline
      \hline
      \noalign{\smallskip}
      Time [BJD$_{\rm TDB}$] & RV~[m~s$^{-1}$] & Unc.~[m~s$^{-1}$]  \\
      \noalign{\smallskip}
      \hline
      $  2455436.50918$  &  $    108.47$  &  $   6.02$  \\
  $  2455436.51698$  &  $     97.80$  &  $   6.86$  \\
  $  2455436.52487$  &  $     88.78$  &  $   6.66$  \\
  $  2455436.53284$  &  $     98.31$  &  $   6.28$  \\
  $  2455436.54061$  &  $    101.71$  &  $   6.31$  \\
  $  2455436.54849$  &  $     77.67$  &  $   5.68$  \\
  $  2455436.56752$  &  $     95.27$  &  $   5.95$  \\
  $  2455436.60407$  &  $     63.98$  &  $   5.07$  \\
  $  2455436.61195$  &  $     51.13$  &  $   5.92$  \\
  $  2455436.61959$  &  $     55.06$  &  $   5.16$  \\
  $  2455436.62746$  &  $     56.87$  &  $   5.15$  \\
  $  2455436.63541$  &  $     66.72$  &  $   4.97$  \\
  $  2455436.64327$  &  $     39.04$  &  $   5.03$  \\
  $  2455436.65130$  &  $     51.37$  &  $   4.76$  \\
  $  2455436.65906$  &  $     16.99$  &  $   5.37$  \\
  $  2455436.66704$  &  $    -35.49$  &  $   5.24$  \\
  $  2455436.67499$  &  $    -10.86$  &  $   4.60$  \\
  $  2455436.68278$  &  $    -33.56$  &  $   5.03$  \\
  $  2455436.69064$  &  $    -41.08$  &  $   5.37$  \\
  $  2455436.69864$  &  $    -56.32$  &  $   5.84$  \\
  $  2455436.70650$  &  $    -46.43$  &  $   5.79$  \\
  $  2455436.71432$  &  $    -58.03$  &  $   5.09$  \\
  $  2455436.72230$  &  $    -65.73$  &  $   5.53$  \\
  $  2455436.73005$  &  $    -55.67$  &  $   5.25$  \\
  $  2455436.73790$  &  $    -69.74$  &  $   5.35$  \\
  $  2455436.74578$  &  $    -58.04$  &  $   4.88$  \\
  $  2455436.75369$  &  $    -51.82$  &  $   5.56$  \\
  $  2455436.76152$  &  $    -58.06$  &  $   5.81$  \\
  $  2455436.76939$  &  $    -61.90$  &  $   6.08$  \\
  $  2455436.77732$  &  $    -55.81$  &  $   6.02$  \\
  $  2455436.78511$  &  $    -17.29$  &  $   5.70$  \\
  $  2455436.79311$  &  $    -30.23$  &  $   5.88$  \\
  $  2455436.80081$  &  $    -11.96$  &  $   6.73$  \\
  $  2455436.80882$  &  $     20.31$  &  $   6.76$  \\
  $  2455436.81669$  &  $     11.38$  &  $   6.45$  \\
  $  2455436.82451$  &  $      0.00$  &  $   6.81$  \\
  $  2455436.83244$  &  $     28.73$  &  $   7.09$  \\
      \noalign{\smallskip}
      \hline
    \end{tabular}  
  \end{center}
\end{table}

\begin{figure*}
  \begin{center}
  \includegraphics[width=17.cm]{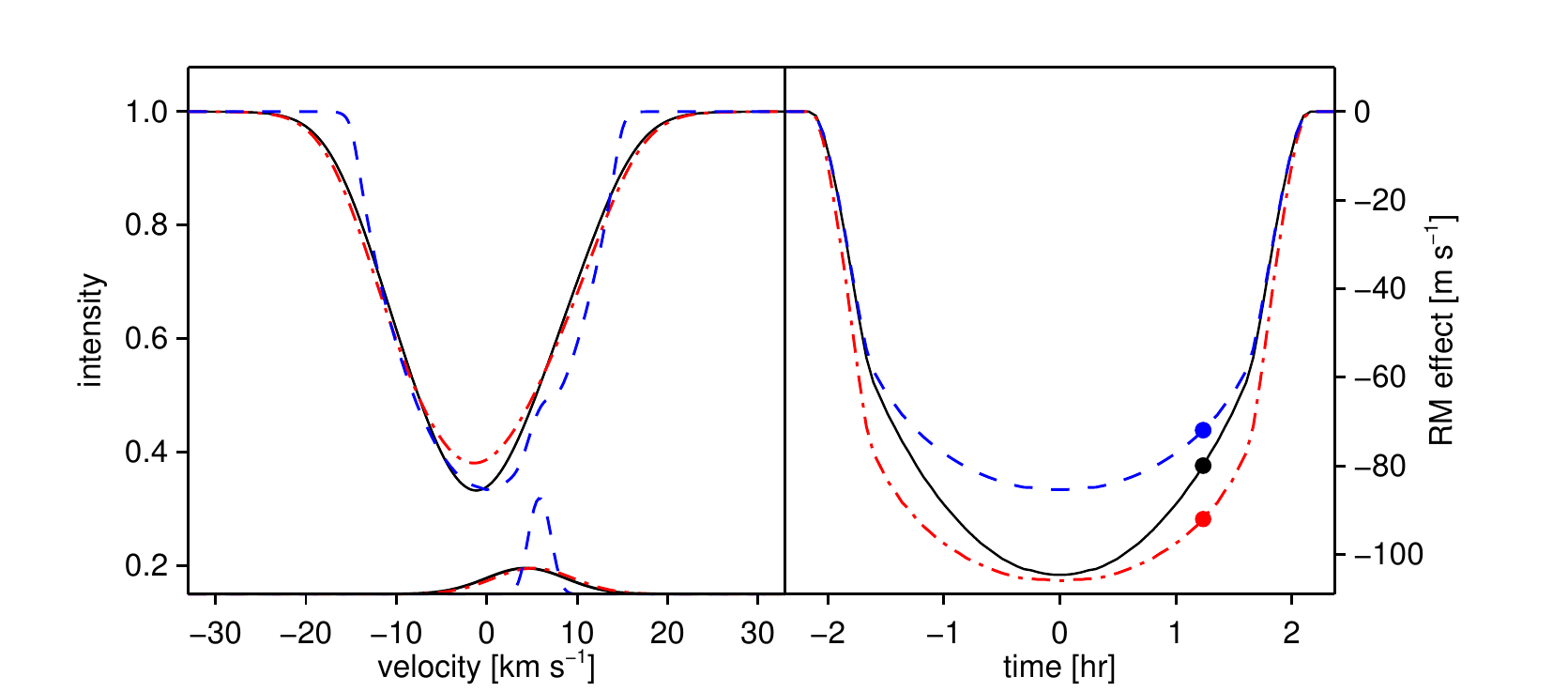}
  \caption {\label{fig:rm_measurement} {\bf Line broadening mechanisms
      and their influence on the RM effect.}  {\it Left}.---Model
    absorption line exhibiting the RM effect. The blue dashed line
    shows the idealized case in which the only line broadening
    mechanism is uniform rotation. The red dash-dotted line shows the
    case in which macroturbulence and the convective blueshift are
    also taken into account. The black line shows the case in which
    the former effects as well as differential rotation are taken into
    account. The lower set of lines indicate the ``loss of light'' in
    the stellar absorption line due to shadowing by the planet. For
    visibility this was multiplied by a factor of 3. These models are
    based on the parameters given in Table~\ref{tab:wasp7_results}.
    {\it Right}.---Time variation of the RM effect (anomalous RV) for
    the same models as in the left panel. The circles indicate the
    particular phase for which the model line profiles in the left
    panel are shown. While the shift in the first moment appears to be
    larger for the blue dashed line (left panel) than for the other
    lines, its RM effect is smaller (right panel). This is partly a
    result of the measurement technique as described in
    section~\ref{sec:rvs}. In addition, the black and red dash-dotted
    lines are influenced by the convective blueshift
    (section~\ref{sec:cb}).}
  \end{center}
\end{figure*}

\subsection{Qualitative expectations}
\label{sec:qualitative}

The RM effect is evident in Figure~\ref{fig:wasp7_rv} as the large
negative velocity excursion (blueshift) that was observed throughout
the transit.  The effect was observed with a high signal-to-noise
ratio.  Simply from the observation that the effect is a blueshift
throughout the transit, we may obtain some information on the
spin-orbit alignment of the system. A qualitative discussion will help
in understanding the quantitative analysis to be discussed later in
this paper.

If the projections of the stellar and orbital spins were aligned, then
the planet would first traverse the half of the stellar surface for which
the rotation velocity has a component directed toward the observer
(blueshifted). The blockage of a portion of this blueshifted half of
the star would cause the absorption lines to appear slightly
redshifted. Then, in the second half of the transit, the reverse would
be true: the anomalous RV would be a blueshift.  In contrast, the RM
effect was observed to be a blueshift throughout the transit. This
implies that the planet's trajectory is entirely over the redshifted
half of the star.

The symmetry of the RM signal about the midtransit time provides
further information. For a star in uniform rotation, the line-of-sight
component of the rotation velocity at any point on the photosphere is
proportional to the distance from the projected rotation axis; see,
e.g., p.~461-462 of \cite{gray2005}.  Therefore if the planet's trajectory
is at a constant distance from the projected rotation axis (i.e.\ if
the projected orbital and spin axes are perpendicular), the hidden
velocity component will be nearly constant in time and the only
variation in the RM signal arises from limb darkening, which is
symmetric about the midtransit time.  In fact this is the only way to
produce a time-symmetric RM effect unless the impact parameter is
nearly zero, which is not the case for WASP-7 \citep{southworth2011}.

We conclude that the projected orbital and spin axes are nearly
perpendicular, and since the anomalous RV is a blueshift we expect
$\lambda\approx 90^\circ$ (as opposed to $-90^\circ$), using the
coordinate system of \cite{ohta2005}.  This qualititative
conclusion is confirmed by our quantitative analysis discussed below;
but first we must take into account several effects besides rotation
that influence the shape of the measured absorption lines.

\subsection{Line broadening and the impact on the analysis}
\label{sec:broaden}

Stellar absorption lines are broadened not only by the Doppler shifts
due to stellar rotation, but also by random motions of material on the
visible stellar surface. These random motions are often referred to as
microturbulence and macroturbulence, depending on whether the length
scale of the velocity field is smaller or larger than the length scale
over which the optical depth is of order unity \cite[see,
e.g.][]{gray2005}.  In this picture, the influence of microturbulence
can be described by a simple convolution by a Gaussian function of an
appropriate width with the rotationally-broadened stellar lines. The
effect of the macroturbulence depends on the angle between the line of
sight and the local normal to the stellar surface. Near the center of
the stellar disk, the Doppler shifts are produced mainly by motions
that are radial with respect to the stellar center; whereas near the
stellar limb, the Doppler shifts are produced by motions tangential to
the stellar surface.  Therefore modeling this effect requires a
spatial integration over the visible hemisphere.  For this purpose we
employ the semi-analytical approach of \cite{hirano2011a}.

Other contributions to the line width come from collisional
broadening, and Zeeman splitting. These effects are small for WASP-7
compared to the other effects, and we describe them by the convolution
of the disk-integrated line profiles with a Lorentzian function of
width $1$~km\,s$^{-1}$.

\subsection{Convective blueshift}
\label{sec:cb}

Another potentially relevant effect on line profiles is the convective
blueshift. At the top of a convective cell, where it meets the
photosphere, the rising material is hotter and therefore more luminous
than the sinking material. Due to the dominance of the rising material
over the falling material, an observer placed vertically above the
stellar surface would detect an overall net blueshift in the
integrated light.  This effectively outward radial velocity is
referred to as the convective blueshift. When integrating over the
visible stellar disk, the convective blueshift from the center of the
disk has a maximal effect on the observed radial velocity, while the
convective blueshift from points near the stellar limb have a weaker
effect because the outward radial velocity has only a small component
along the line of sight. Thus, the influence of the convective blueshift
is strongest near the center of the line profile, and weaker in the
wings of the lines where a substantial portion of the light originates
from the stellar limb.

\begin{figure}
  \begin{center}
   \includegraphics[width=8cm]{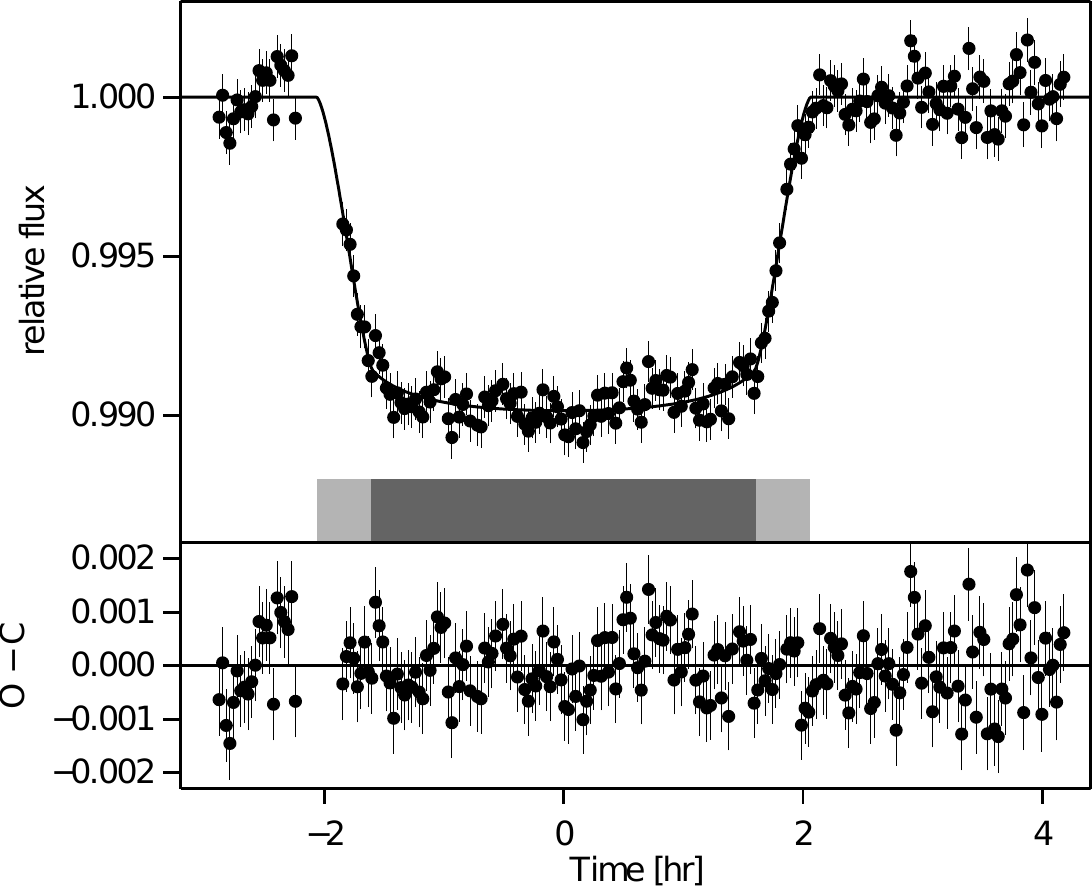}
   \caption {\label{fig:wasp7_phot} {\bf Photometry of WASP-7
       transits.} The upper panel shows the light curve obtained by
     \cite{southworth2011} in the Gunn {\it I} filter, and our
     best-fitting model. The lower panel shows the residuals between
     the data and best-fitting model.}
  \end{center}
\end{figure}

Therefore, the disk-integrated light has not only an
overall Doppler shift but also an asymmetry in the stellar absorption
lines. To describe this, we employ the model of
\cite{shporer2011}. This model captures the
first-order effects of the convective blueshift but ignores a
higher-order asymmetry in the absorption lines: the line cores are
formed relatively high in the photosphere where turbulent motions may
be less important. As WASP-7 is a relatively hot star, with high
turbulent velocities, it seems appropriate to ignore the latter effect
\cite[see Figure 17.15 in][]{gray2005}.

\subsection{Differential rotation}
\label{sec:diff}

When analyzing the RM effect researchers normally assume the eclipsed
star is rotating uniformly (no differential rotation). In the case of
WASP-7, the planet's trajectory is evidently perpendicular to the
projected stellar equator (see section~\ref{sec:qualitative}), and
therefore spans a wide range of stellar latitudes, and the effects of
differential rotation might be expected to be especially important.
Ignoring the possibility of differential rotation might lead to a
systematic error in the measurement of the projected obliquity. For a
well-aligned system ($\lambda \approx 0^\circ$) the effect is much
less important because the planet probes only a small range of stellar
latitudes over the course of the transit \citep[see
also][]{gaudi2007}.

We have limited empirical knowledge of differential rotation profiles
in stars other than the Sun, and therefore we have limited ability to
predict the impact of differential rotation on our observation of
WASP-7. One relevant study is by \cite{reiners2003}, who examined
differential rotation in slowly-rotating stars ($v \sin i_{\star}
\lesssim 20$\,km\,s$^{-1}$). In their sample, all stars having a $log
R'_{HK}$ index between $-4.80$ and $−4.65$ showed signs of differential
rotation, while more active stars had no discernible differential
rotation. Their sample did not include any quieter stars ($log
R'_{HK}<-4.80$).  In this context, we should expect significant
differential rotation for WASP-7, which rotates more slowly then
$20$\,km\,s$^{-1}$ \citep{hellier2009}, and has a low activity index
$log R'_{HK}$ of $-4.981$ (H.\ Knutson, priv.\ communication, 2011).
We adopt the parameterization used by \cite{reiners2003}
and others,
 \begin{equation}
  \label{diffrot}
  \Omega(l) = \Omega_{\rm eq} (1 - \alpha \sin^2{l}).
\end{equation}
Here $l$ denotes stellar latitude and $\Omega_{\rm eq}$ the angular
rotation speed at the equator.  $\alpha$ is a dimensionless parameter
which describes the degree of differential rotation. An $\alpha$ of
$0$ would indicates uniform rotation and an $\alpha$ of $0.2$
corresponds to Sun-like differential rotation. See also
\cite{hirano2011b}, who used the RM effect to set an upper limit on
the stellar differential rotation in the XO-3 system.

\begin{figure}
  \begin{center}
  \includegraphics[width=8.cm]{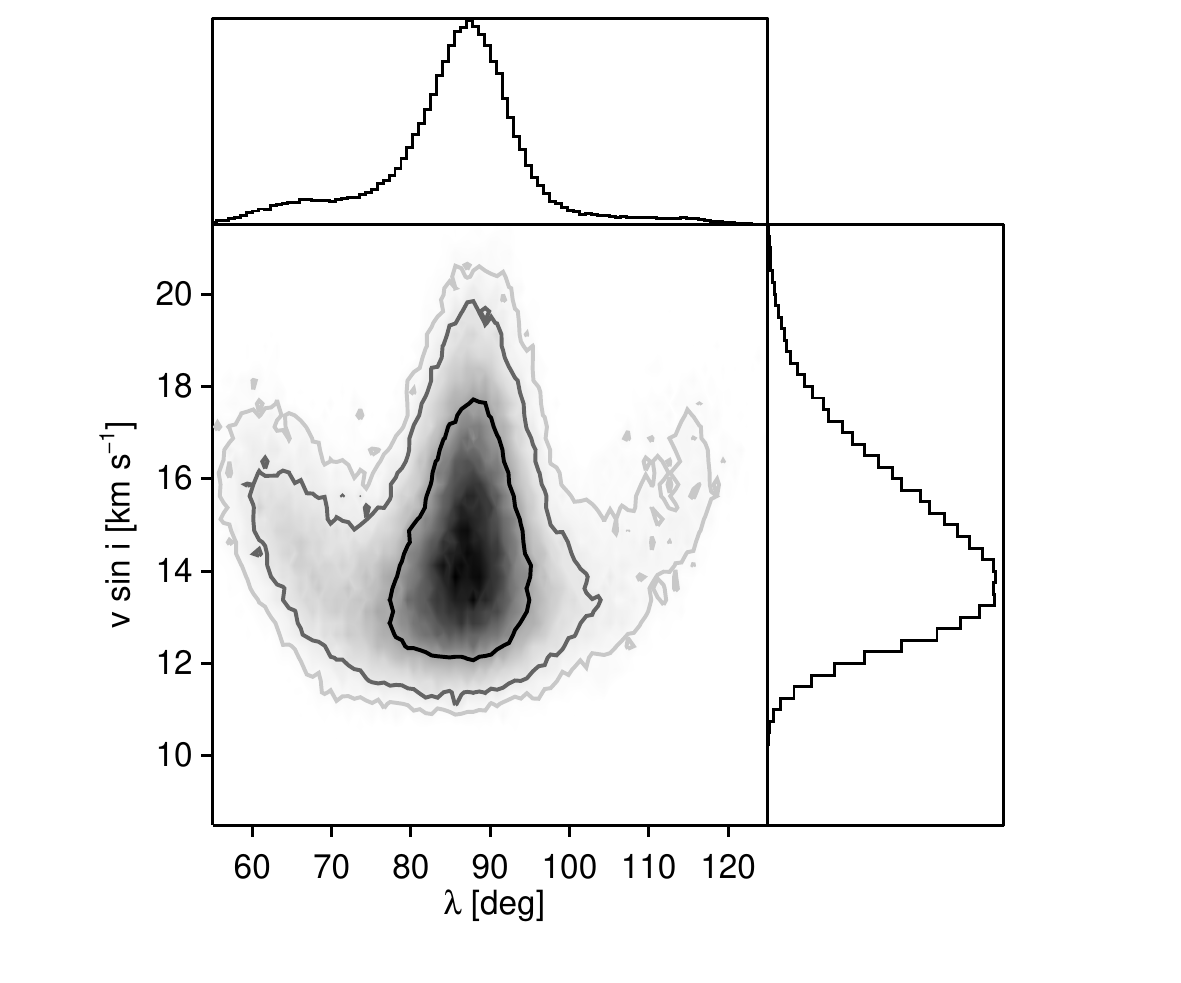}
   \caption {\label{fig:wasp7_mcmc}{\bf Results for $v \sin i_{\star}$
     and $\lambda$, based on our MCMC analysis in the WASP-7 system. }
   The gray scale indicates the posterior probability density,
   marginalized over all other parameters. The contours represent the
   2-D 68.3\%, 95\%, and 99.73\% confidence limits. The
   one-dimensional marginalized distributions are shown on the sides
   of the contour plot. 
}
  \end{center}
\end{figure}

\subsection{Modeling the anomalous RV}
\label{sec:rvs}

In order to develop a model for the anomalous RV produced by the RM
effect, we must take into account two other effects: (1) the stellar
absorption lines recorded by the spectrograph are further broadened by
the PSF of the spectrograph; and (2) the measured RVs represent the
output of a Doppler-measuring code, which is akin to finding the peak
of a cross-correlation between a stellar template spectrum obtained
outside the transit, and a spectrum observed during the transit.  Some
researchers neglect the second point, and model the anomalous RV as
the shift in the first moment of the model absorption line rather than
as the shift in the peak of the cross-correlation function.  Depending
on the system parameters this can lead to systematic errors in the
derived parameters \citep{hirano2010,hirano2011a}.

\begin{table}[t]
 \begin{center}
  \caption{Parameters of the WASP-7 system\label{tab:wasp7_results}}
    \smallskip 
       \begin{tabular}{l  r@{\,\,$\pm$\,\,}l    }
          \tableline\tableline
          \hline
	  \noalign{\smallskip}
	  Parameter &  \multicolumn{2}{c}{ Values} \\
	  \noalign{\smallskip}	 
	  \hline
	  \noalign{\smallskip}
          \multicolumn{3}{c}{Parameters mainly derived from photometry} \\
          \noalign{\smallskip}
          \hline
	  \noalign{\smallskip}
          Midtransit time $T_{\rm c}$ [BJD$_{\rm TDB}$$-$2\,400\,000] &  $55446.635$ & $0.0003$     \\
          Period, $P$ [days]                                                                  &  $4.9546416$   &   $0.0000035$   \\
          $\cos i_{\rm o}$                                                                      &     $0.05$    &  $0.02$ \\   
          Fractional stellar radius,         $R_{\rm \star}/a $                      &  $0.109$  & $^{0.011}_{0.006}$   \\
          Fractional planetary radius,	 $  R_{\rm p}/R_{\star} $              &  $0.096$ &  $0.001$  \\
          $u_{1}$+$u_{2}$                                                                        &  $0.34$ &  $0.08$ \\         
         \noalign{\smallskip}
          \hline
          \noalign{\smallskip}
          \multicolumn{3}{c}{Parameters mainly derived from RVs} \\
          \noalign{\smallskip}
          \hline 
          \noalign{\smallskip}
          Velocity offset, [m\,s$^{-1}$]                                           & $34$ & $5$ \\
          Linear slope during transit night, $a_1$ [m\,s$^{-1}$]                 &   $-95$ & $41$ \\
          Quadratic slope during transit night, $a_2$ [m\,s$^{-1}$/day]                 &   $500$ & $310$  \\
          $\sqrt{v \sin i_{\star}} \sin \lambda$ [km\,s$^{-1}$]                 &  $3.7$ & $0.3$ \\
          $\sqrt{v \sin i_{\star}} \cos \lambda$ [km\,s$^{-1}$]               & $0.24$ & $0.56$  \\
          Macro turbulence parameter, $\zeta$  [km\,s$^{-1}$]             &  $6.4$ &  $1.$ \\      
          $u_{1 rm}$+$u_{2 rm}$                                                                        &  $0.92$ &  $0.06$ \\         
          differential rotation parameter, $  \alpha $                            &  $0.45$ &  $0.11$  \\
          $\cos i_{\star}$                                                                      &     $0.18$    &  $0.43$ \\   
          \noalign{\smallskip}
          \hline
          \noalign{\smallskip}
          \multicolumn{3}{c}{Indirectly derived parameters} \\
	  \noalign{\smallskip}
          \hline
          \noalign{\smallskip}
          Orbital inclination,   $i_{\rm o}$   [$^{\circ}$]             &       $87.2$ &  $^{0.9}_{1.2}$    \\
          Full duration, $T_{14}$  [hr]           &     $4.12$ &  $^{0.09}_{0.06}$     \\
          Ingress or egress duration, $T_{12}$  [min]      &  $27$ &    $^{6}_{9}$       \\
          Projected stellar rotation speed, $v \sin i_{\star}$ [km\,s$^{-1}$]     &   $14$ & $2$            \\
          Projected spin-orbit angle,  $\lambda$    [$^{\circ}$]             &      $86$ & $6$ \\
	  \noalign{\smallskip}
	  \tableline
           \noalign{\smallskip}
           \noalign{\smallskip}
     \end{tabular}
   \end{center}
\end{table}

\cite{hirano2011a} developed an analytical description for the shift in
the cross-correlation peak as a function of the transit parameters,
the stellar rotation velocity and obliquity, the microturbulent and
macroturbulent velocities, the differential rotation profile, and the
PSF width of the spectrograph. We use their approach to model the
obtained RVs during transit and also include a model as described
above for the convective blueshift. See
Figure~\ref{fig:rm_measurement} for an illustration of how the above
described effects change the line shape and the time variation of the
RM effect.

\subsection{Other RV variation sources}
\label{sec:rv_var}

To this point we have only discussed changes in the RVs due to the
transit. To successfully model the RM effect we also need to model the
change in radial velocity due to the orbital motion of the star.  In
addition, as we will see in section~\ref{sec:jitter_ecc}, WASP-7 shows a high
level of RV noise on a timescale of days, which will introduce trends
in the RV over the course of the transit night. Therefore our model
allows for RV trends that are linear or quadratic functions of time,
in addition to the RM effect and orbital motion.  The physical
interpretation of these trends is discussed in section
\ref{sec:jitter_ecc}.

\subsection{Quantitative Analysis}
\label{sec:quantitative}

Now that all the ingredients of our model have been introduced, we
describe the various parameters in detail. The Keplerian orbital
motion of the star is specified by the period ($P$), the time of
inferior conjunction ($T_{\rm c}$), the semi-amplitude of the
projected stellar reflex motion ($K_{\star}$), and a velocity offset
($\gamma$). Initially we assume the orbit to be circular, as
\cite{hellier2009} found no sign of an orbital eccentricity (although
see Section~\ref{sec:jitter_ecc}). We also allow for additional linear
and quadratic RV trends ($a_1$ and $a_2$) on the transit night,
\begin{equation}
  RV(t) = \gamma + RV_{\rm Orbit}(t) +RV_{\rm RM}(t) + a_1*(t-t_{0}) + a_{2}*(t-t_{0})^2\,\,\,.
\end{equation}
Here RV$_{\rm Orbit}(t)$ and RV$_{\rm RM}(t)$ represent the radial
velocities caused by the orbital motion and the RM effect. $t_{0}$ is
a point in time near the middle of our observation sequence.
Since allowing for these trends we have relinquished any power to
constrain $K_{\star}$ with the data, so we fix $K_{\star}$ at the
value reported by \cite{hellier2009}.

The amount of light blocked at any given phase of the transit depends
on the location of the chord of the planet's path over the stellar
disk, which is parameterized by the cosine of the orbital inclination
($\cos i_{\rm o}$), the radius of the star in units of the orbital
semimajor axis ($R_{\star}/a$), and the radius of the planet relative
to the stellar radius ($R_{\rm p}/R_{\star}$). We use quadratic limb
darkening parameters $u_{1 rm}$ and $u_{2 rm}$ to parameterize the
relative intensity of the stellar surface in the wavelength region of
$5000$-$6200$\,\AA, the spectral region in which the RVs are
measured. We chose limb-darkening coefficients $u_{1 rm}=0.3$, $u_{2
  rm}=0.35$, based on the tables of \cite{claret2004}. We allowed
$u_{1 rm}+u_{2 rm}$ to vary freely and held fixed $u_{1 rm}-u_{2 rm}$
at the tabulated value of $-0.05$, since the difference is only weakly
constrained by the data (and in turn has little effect on the other
parameters).

We parametrize the projected spin-orbit angle and the projected
stellar rotation by the quantities $\sqrt{v\sin i_\star} \cos\lambda$
and $\sqrt{v\sin i_\star} \sin\lambda$, rather than $v\sin i_\star$
and $\lambda$. We do this because our chosen parameters are less
strongly correlated \citep[e.g.][]{albrecht2011b}. Our macroturbulence
model was that of \cite{gray2005}, assuming equal surface fractions of
radial and tangential velocities, with a macroturbulence parameter
$\zeta$ subject to a Gaussian prior of $6.4\pm1.0$~km\,s$^{-1}$. This
represents our expectation for a star of WASP-7's effective
temperature \citep{gray1984}. To model the convective blueshift we
assumed an outward blueshift of $1$~km\,s$^{-1}$ at all positions on
the star \citep[][and references therein]{shporer2011}. We further
include $\alpha$, the parameter which governs the strength of
differential rotation, and the stellar inclination towards the
observer ($i_{\star}$) as free parameters. As \cite{reiners2003} found
no star with $\alpha \sqrt{\sin {i_{\star}}}$ greater $\approx0.4$
(see their Figure~16) we also impose a prior on $\alpha$ which is flat
until $0.4$ and then falls off as a Gaussian function with a standard
deviation of $0.1$.

Finally, we must specify the width of a Gaussian function representing the
width of the lines due to both microturbulence and the PSF of the
spectrograph. We chose $\sigma=3$~km\,s$^{-1}$ for this purpose. Also,
to represent the natural broadening of the lines we used a Lorentzian
function with a width of $1$~km\,s$^{-1}$ \cite[see
also][]{hirano2011a}.

Additional information on the transit geometry comes from a transit
light curve recently obtained by \cite{southworth2011}. They made
their de-trended light curve available via
\href{http://vizier.cfa.harvard.edu/viz-bin/VizieR}{VizieR}. They also
gave an updated transit ephemeris for the system, which is derived from the WASP
discovery data in combination with the new light curve. We used
those updated results for $P$ and $T_{\rm c}$ as priors (see Equation
\ref{equ:wasp7_mcmc}). We also fitted the light curve simultaneously
with the RVs in order to pin down $\cos i_{\rm o}$, $R_\star/a$, and
$R_p/R_\star$.  We used the algorithm from \cite{mandel2002} to model
the light curve which was obtained in the Gunn {\it I} filter. From
\cite{claret2004} we obtained $u_{1}=0.17$ and $u_{2}=0.36$. We allowed
$u_{1}+u_{2}$ to vary freely, and held $u_{1}-u_{2}$ fixed at the
tabulated value of $-0.19$.

To derive confidence intervals for the parameters we used the Markov
Chain Monte Carlo algorithm.  The likelihood was taken to be
$\exp(−\chi^{2} /2)$, where $\chi^{2}$ was defined as: {\small
 \begin{eqnarray}
\label{equ:wasp7_mcmc}
\chi^{2} & = &
     \sum_{i=1}^{57} \left[\frac{{\rm RV}_i{\rm (o)}-{\rm RV}_i{\rm (c)}}{\sigma_{{\rm RV}, i}}\right]^2 +
   \sum_{j=1}^{1134} \left[\frac{{\rm F}_j{\rm (o)}-{\rm F}_j{\rm (c)}}{\sigma_{{\rm F}, j}}\right]^2 \nonumber\\
  &  &  + \left(\frac{T_{\rm c, BJD} - 2455446.63493}{0.000030}\right)^2 \nonumber\\
&  & +\left(\frac{P- 4\fd9546416}{0\fd0000035}\right)^2  +
     \left(\frac{\zeta- 6.425\, {\rm km\,s}^{-1}}{ 1\, {\rm km\,s}^{-1}}\right)^2 \nonumber\\
&  & + \left\{ \begin{array}{l l}
     0 & \quad \text{if $\alpha \leq 0.4$}\\
     \left(\frac{\alpha - 0.4}{0.1}\right)^2 & \quad \text{if $\alpha > 0.4$ }\\
   \end{array} \right. ,
\end{eqnarray}
}

\begin{figure}
 \begin{center}
  \includegraphics[width=8.cm]{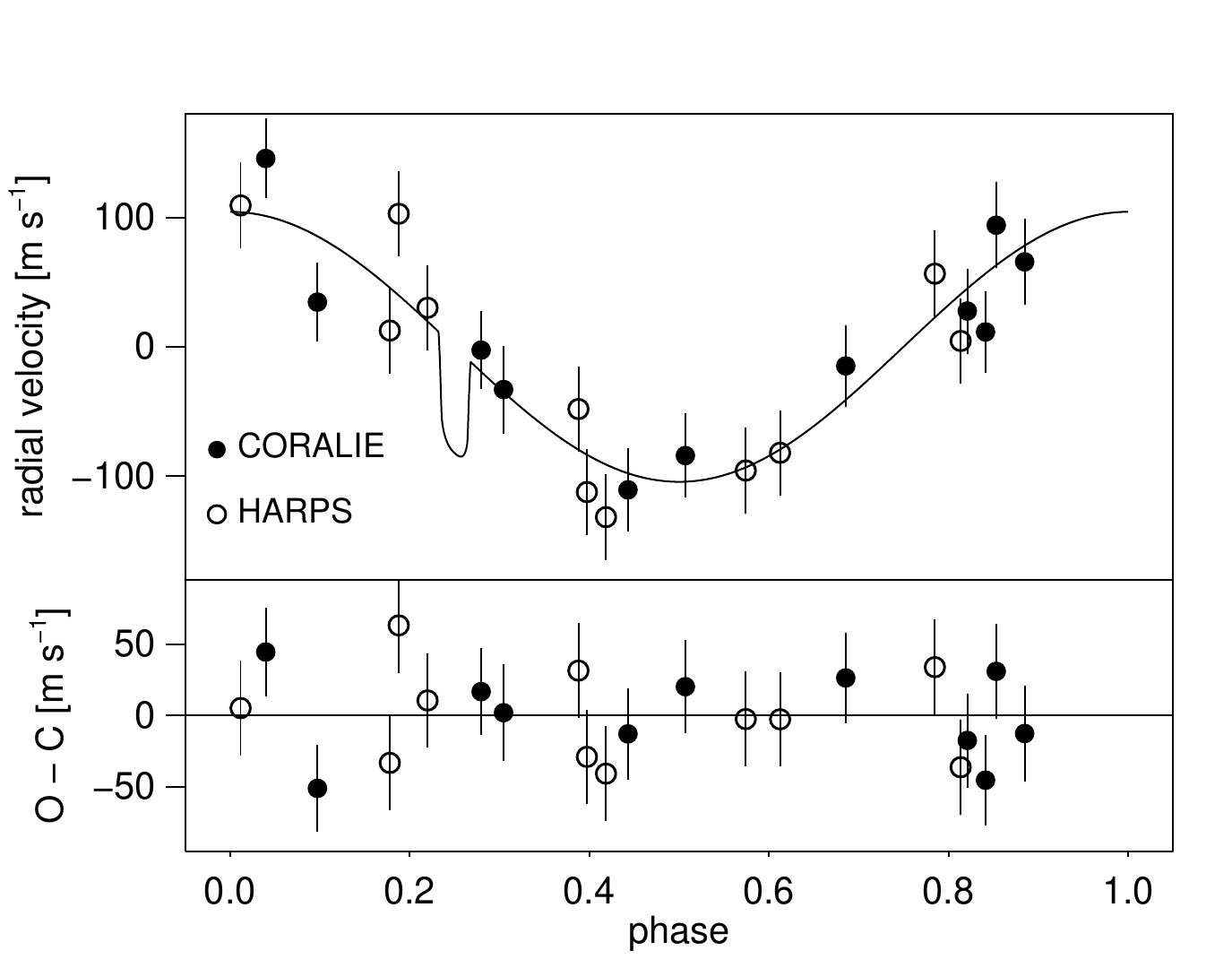}
  \caption {\label{fig:wasp7_ecc_orbit} {\bf Orbital solution with zero eccentricity for
      WASP-7.} The upper panel shows RV observations as function of
    the orbital phase, with periastron at zero phase. The CORALIE RVs are
    indicated by filled symbols and the HARPS data are shown by open
    symbols. The lower panel shows the residuals between data and
    best-fitting model.}
 \end{center}
\end{figure}

The first two terms are sums-of-squares over the residuals between the
observed (o) and calculated (c) values of the radial velocity (RV) and
relative flux (F). The following terms represent priors on some
parameters, as mentioned above. Before starting the chain we also added
$10$~m\,s$^{-1}$ in quadrature to the uncertainty of the PFS RVs to
obtain a reduced $\chi^{2}$ close to unity. In making this step we
assumed that the uncertainties in the RV measurements are uncorrelated
and Gaussian.

Our results are presented in Table~\ref{tab:wasp7_results}. The best
fits to the RVs and photometry are shown in Figures~\ref{fig:wasp7_rv}
and \ref{fig:wasp7_phot}. Figure~\ref{fig:wasp7_mcmc} shows the 2-d
posterior density distribution for the two parameters of greatest
interest for our study: $\lambda$ and $v \sin i_{\star}$.

For the parameters governed mainly by the photometric data, our
results are consistent with those obtained by
\cite{southworth2011}. We will discuss the reliability of some of the
parameters found by the RV data, in particular spectral limb
darkening, the evidence for differential rotation, and the stellar
inclination in section~\ref{sec:diffrot}, after investigating possible
reasons for the excess noise in the RV data in the following section.

\section{Stellar Jitter and orbital eccentricity}
\label{sec:jitter_ecc}

The out-of-transit RV gradient measured with the PFS is steeper than
would be expected from the previously published spectroscopic
orbit. That slope, and the known orbital period, imply an orbital
velocity semiamplitude of $K_{\star} \approx 200$~m\,s$^{-1}$, in
strong contrast to the published value of $97\pm13$~m\,s$^{-1}$. (See
the upper panel in Figure~\ref{fig:wasp7_rv}.)

Archival RV data from the CORALIE and HARPS spectrographs, obtained
during various orbital phases, show a large scatter around the best
orbital solution in excess of the measurement uncertainties. This type
of excess noise is commonly referred to as ``stellar jitter''. In the
following we will shortly investigate the timescale over which the RV
scatter is correlated and test the possibility that the orbit is
actually eccentric.

\subsection{Excess noise}
 
We turn first to the previously reported RVs.  Eleven measurements
were obtained with CORALIE by \cite{hellier2009} and another eleven
RVs were measured with HARPS by \cite{pont2011}. We fitted an orbital
model to both data sets separately to determine the root mean square
(rms) residual between the data and the best-fitting model. A high rms
would indicate that our model neglects an effect which influences the
RVs, which for example could be activity of the star itself. We find
that the scatter of both the CORALIE and HARPS RVs is not only greater
than what these instruments normally achieve on bright stars, also the
rms of the HARPS data ($33$~m\,s$^{-1}$) is slightly higher than the
rms of the CORALIE RVs ($31$~m\,s$^{-1}$). This is noteworthy as HARPS
should achieve a greater precision then CORALIE, if photon noise is
the limiting factor, as HARPS operates in conjunction with a $3.6$\,m
telescope and CORALIE with a $1.2$\,m telescope. This suggests that an
additional source of RV variations may be present, which dominates the
noise budget.

Next we fitted each dataset individually, assuming a circular orbit
and adding a term in quadrature to the internal uncertainties to
produce a reduced $\chi^{2}$ of unity.  In making this step we assumed
that the errors of the RV measurements are uncorrelated and Gaussian.
We needed to add $25$~m\,s$^{-1}$ in quadrature to the internal
uncertainties of the eleven CORALIE datapoints and $33$~m\,s$^{-1}$ to
the uncertainties of the eleven HARPS RVs.

Using those inflated uncertainties, we fitted both data sets together,
also assuming a circular orbit. Our fitting statistic was
{\small
\begin{eqnarray}
\label{equ:orbit_mcmc}
\chi^{2} & = &
     \sum_{i=1}^{22} \left[\frac{{\rm RV}_i{\rm (o)}-{\rm RV}_i{\rm (c)}}{\sigma_{{\rm RV}, i}}\right]^2  + \left(\frac{T_{\rm c, BJD} - 2455446.63493}{0.000030}\right)^2 \nonumber\\
&  & +\left(\frac{P- 4\fd9546416}{0\fd0000035}\right)^2 ,
\end{eqnarray}
}making use of the ephemeris from \cite{southworth2011}.

\begin{figure}
 \begin{center}
  \includegraphics[width=8.cm]{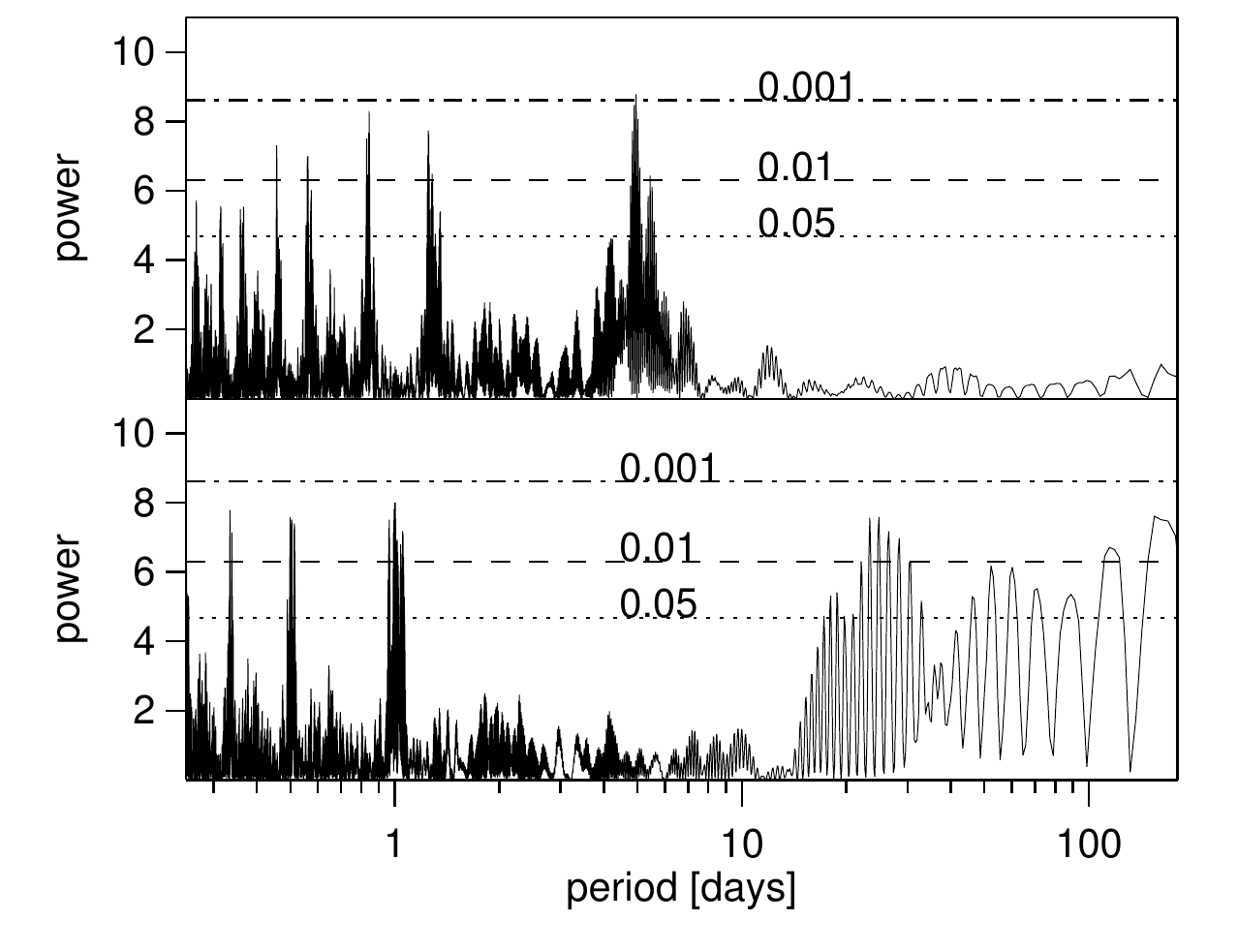}
  \caption {\label{fig:ls} {\bf Lomb-Scargle periodograms of the
      out-of-transit RVs.}
    The upper panel shows the power at different
    periods in the RV data of the two data sets from CORALIE and HARPS.
    The highest peak occurs at the orbital period. The lower panel
    shows the power in the RV data after the best fitting model was
    subtracted. Significant peaks occur at a period of one day, and
    its higher harmonics. Most likely this is a consequence of the
    diurnal time sampling of the RV observations.}
 \end{center}
\end{figure}

Figure~\ref{fig:wasp7_ecc_orbit} shows the phase-folded RV data.  As
expected from the previous fits, the rms residual is $31.4$
~m\,s$^{-1}$.  To understand this noise source it would be important
to learn the timescale over which the RV noise is correlated.  Looking
at the Lomb-Scargle periodogram of the RV data, we find the most
dominant peak at the orbital period of WASP-7b (Figure~\ref{fig:ls},
upper panel). After subtracting the best-fitting circular-orbit model,
the periodogram of the residuals shows a strong peak at a period of
one day and its harmonics (Figure~\ref{fig:ls}, lower panel). This is
most likely a result of the timing of the observations. Many
observations were obtained on consecutive nights. Apart from this
pattern, the periodogram is not very informative. An autocorrelation
plot does not reveal additional information.

The timing is different for the PFS observations. We obtained 37 data
points during an interval of $7.5$\,hr. The rms around our solution
including the orbital model, the linear and quadratic acceleration, and
the RM effect is $11$~m\,s$^{-1}$ (Figure~\ref{fig:wasp7_rv}). This
relatively low scatter indicates that the correlation time of the RV
variations is longer then a few hours. 

We note that WASP-7 is not the only early-type planet host star for
which a large stellar jitter has been observed. Recently
\cite{hartman2011} found that the two planet host stars in the
HAT-P-32 and HAT-P-33 systems have stellar jitter of $\approx 80$ and
$55$~m\,s$^{-1}$.

\begin{figure}
  \begin{center}
  \includegraphics[width=8.cm]{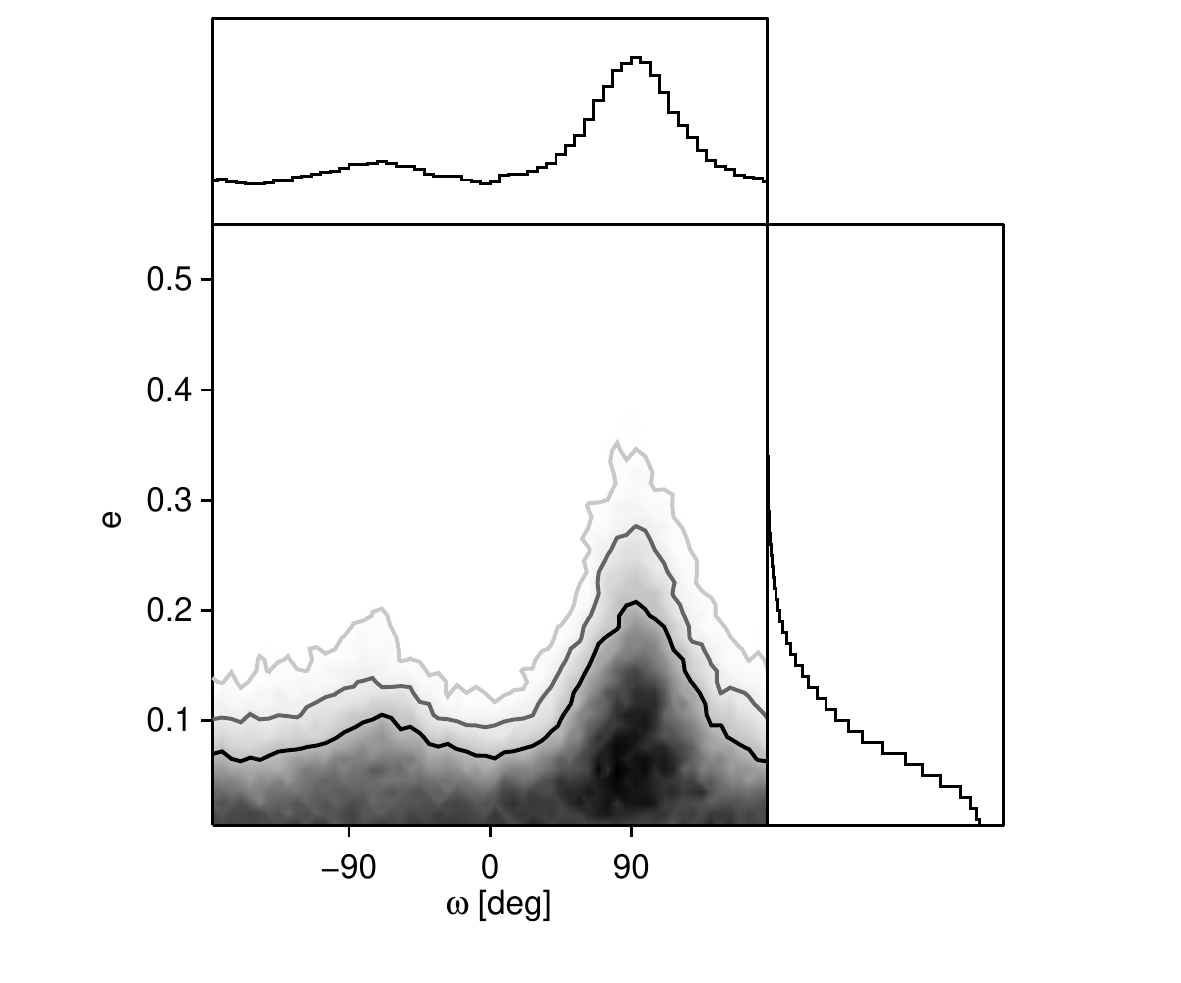}
  \caption {\label{fig:wasp7_omega_ecc} {\bf Constraints on the orbital
      eccentricity ($e$) and orientation ($\omega$) of WASP-7 from
      HARPS and CORALIE data.} The gray scale indicates the posterior
    probability density, marginalized over all other parameters. The
    contours represent the 2-d 68.3\%, 95\%, and 99.73\% confidence
    limits. The eccentricity is constrained to low values. For an
    orientation of $\omega \approx |90|^{\circ}$ somewhat higher
    values of the eccentricity are allowed. That is particularly true
    for positive value of $\omega$. The 1-d marginalized
    distributions are shown on the sides of the contour plot.}
  \end{center}
\end{figure}

\subsection{Orbital eccentricity}

One possible contributing factor to the excess RV noise is that the
orbit is actually eccentric, in violation of our modeling assumption
of a circular orbit. To investigate this possibility we repeated the
MCMC analysis but allowed the orbital eccentricity ($e$) and the
argument of periastron ($\omega$) to be free parameters. Our stepping
parameters were $\sqrt{e} \sin \omega$ and $\sqrt{e}\cos\omega$, which
have less correlated uncertainties for small eccentricity, and which
correspond to a flat prior in $e$.

Using the likelihood based on Equation~\ref{equ:orbit_mcmc}, we
derived the posterior probability, which is displayed in the
$\omega$--$e$ plane in Figure~\ref{fig:wasp7_omega_ecc}. The rms for
the best fitting eccentric solution is with $30.9$~m\,s$^{-1}$, only
moderately smaller than for the circular-orbit model. We can see from
Figure~\ref{fig:wasp7_omega_ecc} that there is no clear detection of
an eccentric orbit and that solutions with $e\gtrsim 0.2$ are
generally disfavored except for values of $\omega$ near $90^\circ$,
for which somewhat larger eccentricities are allowed.  Such an
argument of periastron would indicate for transiting systems a
semi-major axis closely aligned with the line of sight (LOS).
Interestingly, such an orbital orientation would lead to a steeper RV
slope during the transit night, which is indeed what was observed with
the PFS.

However, the peak in the probability density near $\omega\approx
90^\circ$ should not be taken as evidence that the orbit really does
have this orientation. This is probably just the result of the fact
that RV studies are better at constraining $\sqrt{e} \times \cos
\omega$ than $\sqrt{e} \times \sin \omega$, and consequently we expect
the confidence interval for $e$ to be larger for $\omega$ close to
$|90|^{\circ}$. Such an orbital configuration would lead to symmetric
RV curves even for an eccentric orbit, i.e., the RV amplitudes at the
quadratures would not differ.

\begin{figure}[t]
  \begin{center}
  \includegraphics[width=8.cm]{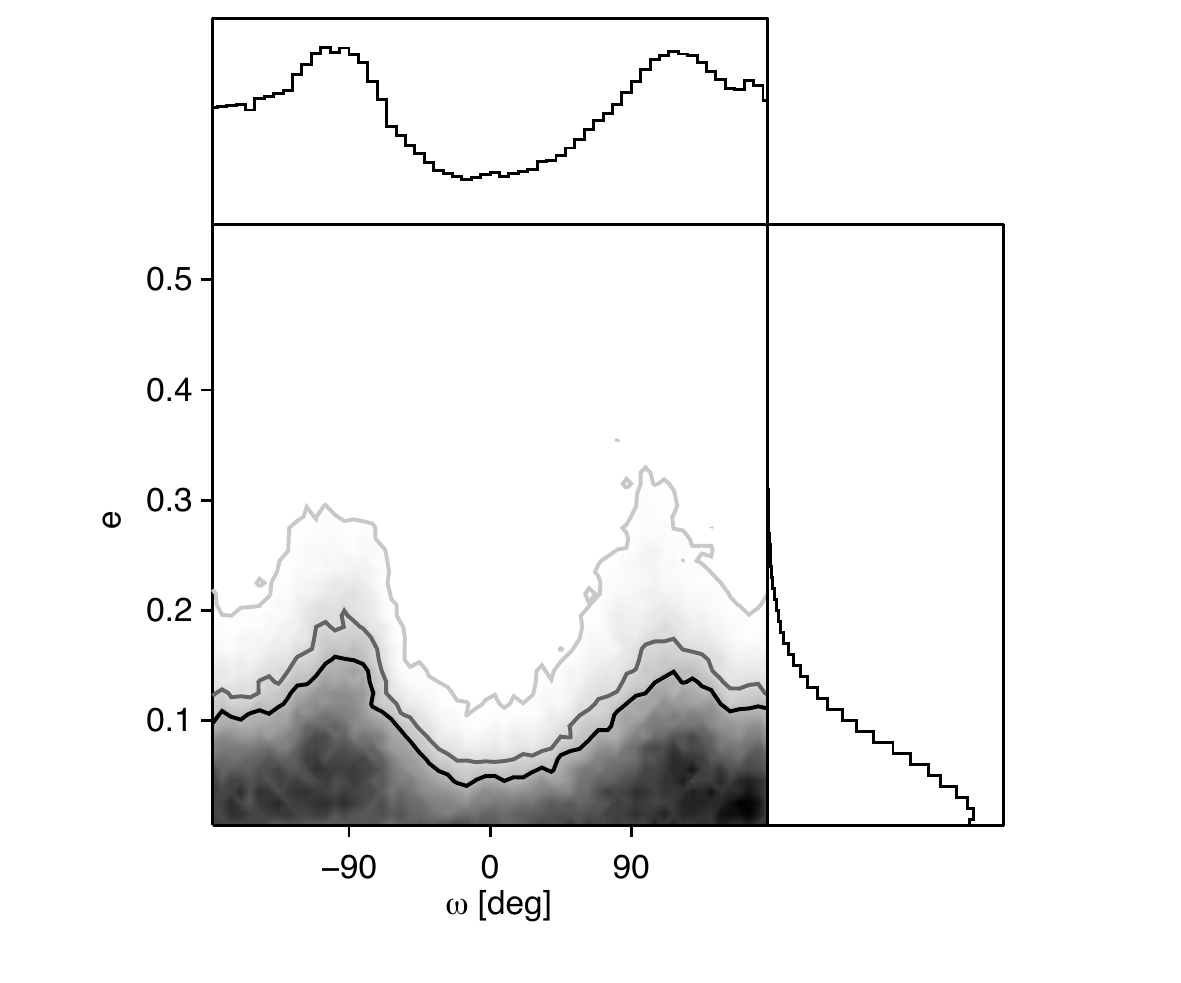}
  \caption {\label{fig:mook_data_omega_ecc} {\bf Results for orbital
      eccentricity using synthetic data.} Similar to
    Figure~\ref{fig:wasp7_omega_ecc} but this time for an MCMC
    analysis of a synthetic data set which was derived from a
    circular-orbit model. The results indicate that higher
    eccentricities are allowed for $\omega$ near $90^\circ$ or
    $-90^\circ$.}
  \end{center}
\end{figure}

To investigate this point further we conducted numerical experiments
similar to those carried out by \cite{laughlin2005} for the HD\,209458
system.  We created simulated RV datasets with the same time stamps as
the CORALIE and HARPS data sets. For this we assumed a circular orbit
with the parameters of the WASP-7 system. We then added Gaussian
perturbations to the model RVs, adopting a 1$\sigma$ uncertainty of
$33$~m\,s$^{-1}$, and then used the same MCMC analysis for these mock
data as was used for the real data sets. A typical 2-d posterior
resulting from this experiment is shown in
Figure~\ref{fig:mook_data_omega_ecc}.

We found that greater values of $e$ are permitted for orbital
orientations of $\approx |90|^{\circ}$. This should make us suspicious
of any low-SNR detection of an orbital eccentricity with an $\omega$
of $\approx |90|^{\circ}$.  We conclude that the out-of-transit RVs
give no indication of an eccentric orbit for WASP-7.  This is in line
with the upper limit $e<0.25$ found by \cite{pont2011}. We note that
the $\omega$-dependent sensitivity to $e$ is very similar to the
$v\sin i_\star$-dependent sensitivity to $\lambda$ that was explicated
by \cite{albrecht2011b}, for low-SNR studies of the RM effect.  In
that case, higher values of the stellar rotation rate ($v\sin i$) are
allowed for $\lambda \approx 0^{\circ}$ and $\approx 180^{\circ}$,
even when the underlying signal has no RM effect at all.

\section{Discussion}
\label{sec:discussion}

\subsection{Differential rotation parameters}
\label{sec:diffrot}

Our model took into account the possible effects of differential
rotation, through the parameters $\alpha$ and $i_\star$.  We found
that the upper boundary of the posterior distribution for $\alpha$ is
determined mainly by our prior. Specifically we found $\alpha=0.45\pm
0.11$, with the prior enforcing $\alpha < 0.5$ (see
Table~\ref{tab:wasp7_results}). If instead no prior is placed on
$\alpha$, then we found that the differential rotation parameter
increases to values as high as $0.9$, much higher than would be
expected based on theory and on observations of other stars.

This should make us suspicious. For $|\lambda|$ close to $90^{\circ}$,
as is the case here, a parameter degeneracy exists between the
limb-darkening profile and the degree of differential rotation,
because both of those phenomena produce changes to the RM effect that
are symmetric in time about the transit midpoint.  Perhaps our
assumptions are mistaken regarding the stellar limb darkening within
the effective observing bandpass.  For RM observations the effective
observing bandpass is complicated to describe, depending as it does
upon the density of $I_2$ absorption lines. Furthermore, there may be
systematic differences between the limb darkening as observed in
different absorption lines, and even within a single strong line, as
different parts of the absorption lines form in different depths of
the stellar photosphere.  Nevertheless, the fitted value of the
center-to-limb variation ($u_{1 rm} + u_{2 rm} = 0.92\pm 0.06$) is
already quite high; lowering this value would only cause $\alpha$ to
converge to {\it larger} values.

The data seem to be demanding a greater variation of the RM effect
between the second and third contact than is delivered by the expected
amounts of limb darkening and differential rotation.  Some possible
explanations are:
\begin{itemize}

\item The stellar jitter exhibits correlations on the timescale of
  minutes to hours, which is not taken into account in our model.  By
  fitting a linear and quadratic trend to the RVs, we have only taken
  into account correlations on longer timescales.

\item If WASP-7 is pulsating then velocity fields on the stellar
  surface, due to the pulsations, might lead to a different shape of
  the stellar absorption line than is created by our model, which does
  not include pulsations. The planet would occult during its transit
  different velocity components than expected. For this effect to be
  important the velocity fields do not need to change on a timescale
  comparable to the transit.

\item Another possibility is that our model of the convective
  blueshift is not correct. For hotter stars there exists a reversed
  shape in the bisectors, although WASP-7 seems to be securely located
  on the 'cool' side of this 'granulation boundary' \citep[see
  Fig.~17.18 in][]{gray2005}.

\end{itemize}

With the present data we cannot determine if one of these or another
effect is responsible for the relatively strong variations in the
shape of the RM signal. One should therefore view our particular
results for the differential rotation, stellar inclination, and the
spectroscopic limb darkening coefficients with some skepticism.
Including these effects, or some effects with a similar functional
form, is however important for a realistic uncertainty estimation of
$v \sin i_{\star}$ and $\lambda$ as we will discuss in the next
section.

\subsection{Stellar obliquity}
\label{sec:obliquity}

As expected from the qualitative discussion we find
$\lambda=86\pm6^{\circ}$, which is consistent with $90^{\circ}$. The
projected stellar spin axis is lying nearly within the orbital plane.

\begin{figure}
  \begin{center}
  \includegraphics[width=8.cm]{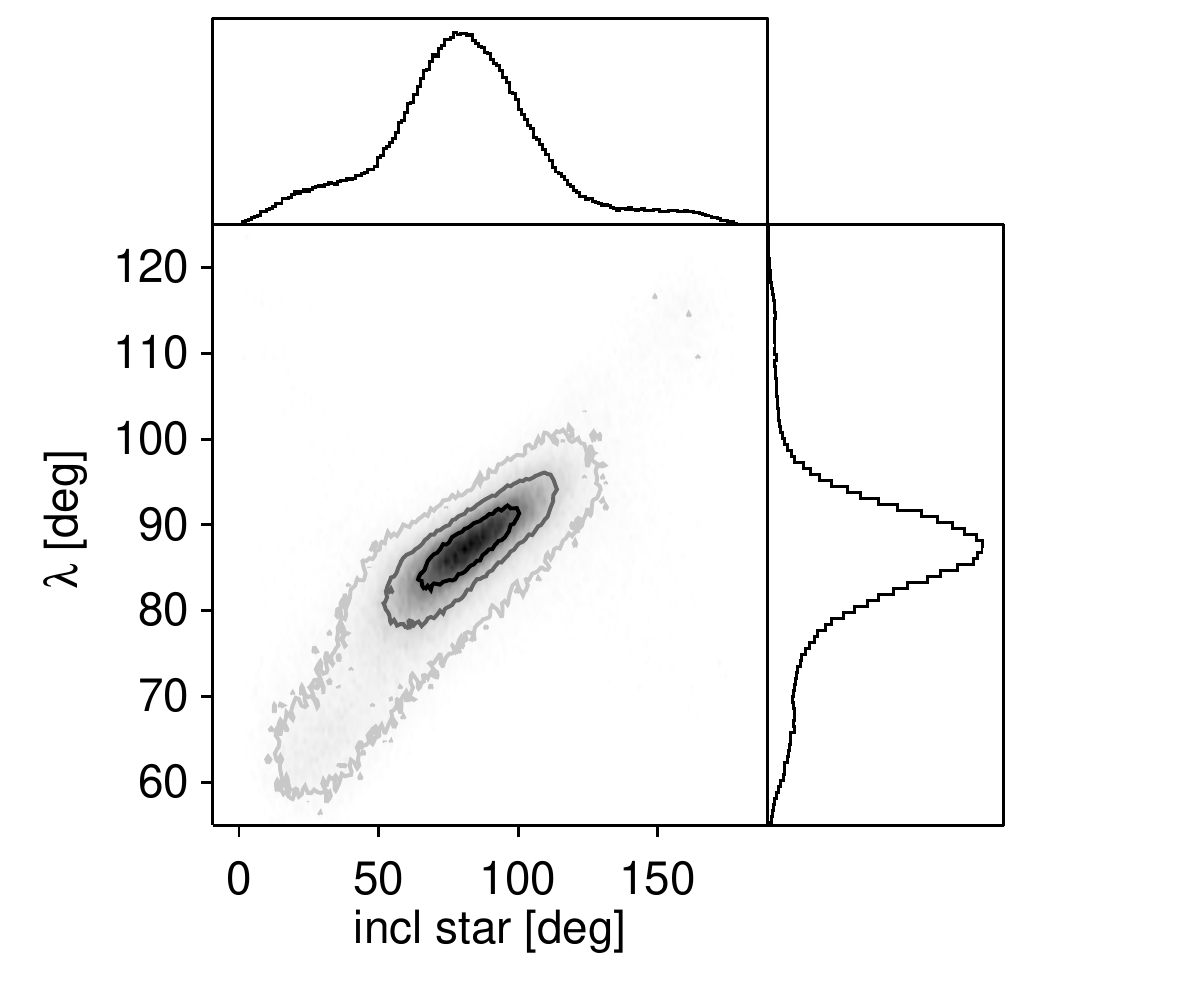}
  \caption {\label{fig:wasp7_incl_proj_obli} {\bf Dependency of
      $\lambda$ on $i_{\star}$.} The gray scale indicates the posterior probability density,
   marginalized over all other parameters. The contours represent the
   2-d 68.3\%, 95\%, and 99.73\% confidence limits. The
   1-d marginalized distributions are shown on the sides
   of the contour plot. One can see the strong correlation between
   $\lambda$ and $i_{\star}$. This
   correlation vanishes if no differential rotation is present.}
  \end{center}
\end{figure}

We did not use the measurement of the projected rotational velocity
($v \sin i_{\star}=17\pm2$~km\,s$^{-1}$) by \cite{hellier2009} as a
prior constraint. This is because it is not clear what value of
macroturbulence was assumed by those authors, and also because if
differential rotation is present then a systematic error could be
introduced.

The lack of knowledge on differential rotation and the stellar
inclination did lead to an increased confidence interval for
$\lambda$. This can be seen from the Figures \ref{fig:wasp7_mcmc} and
\ref{fig:wasp7_incl_proj_obli}. When the stellar inclination departs
from $90^{\circ}$, the planet covers higher stellar latitudes,
which have decreased rotational velocities (and therefore a decreased
RM effect) either at the beginning or end of the transit. This forces
a higher $v \sin i_{\star}$ and a change in $\lambda$ to compensate
for the asymmetry in the transit RV curve.  This degeneracy could be
broken if we would have independent information on the stellar
inclination.

To estimate $\sin i_\star$ we can use the technique of
\cite{schlaufman2010}, involving a comparison of the measured value of
$v\sin i_{\star}$ with the expected value of $v$ for a star of the
given age and mass. Schlaufman found for WASP-7 a rotation statistic
$\Theta = -3.4$, indicating that WASP-7 rotates faster then expected
for its age and mass. A $\Theta$ near $0$ would indicate that the
measured $v\sin i_{\star}$ is consistent with the expected rotation
speed $v$ for a star of a given mass and age. A $\Theta$ larger then
$0$ would indicate an inclination of the stellar spin axis towards the
observer.

We repeat his analysis with the new values for $v\sin i_{\star}$,
found here using the RM effect (Table~\ref{tab:wasp7_results}), and
the new mass, radius and age values from \cite{southworth2011}. With
these we obtain an expected $v$ of $11\pm4$~km\,s$^{-1}$ and a
rotation statistic $\Theta$ of $-0.8$. This indicates that the
projected rotation speed is consistent with a $\sin i_{\star}$ of
$\approx 1$ There is no indication of an inclination of the stellar
spin axis towards the observer.

This knowledge could be used as prior knowledge, or more correctly
used in the model itself to decrease the uncertainty in the projected
obliquity. However as WASP-7 is at the upper end of the mass range for
which \cite{schlaufman2010} calculated his rotation, mass, age
relationship and because of the complications due to differential
rotation for which his relation was not calibrated, we decided not to
use this knowledge to reduce the uncertainty in $\lambda$.

\subsection{Stellar jitter}
\label{sec:jitter}

Analyzing the bisectors of the obtained spectra might lead to a
reduced scatter, as was done for HAT-P-33 by \cite{hartman2011}. In
particular the data taken during the transit night might be
informative as the change in the bisectors will be correlated over the
course of the night.

If during some nights several RVs would have been obtained then the
uncertainty in the orbital parameters of WASP-7 could be reduced in a
similar approach to that used by \cite{hatzes2010} for CoRoT-7. They used
data obtained during one night to constrain a part of the orbit and
allowed for a drift between different nights. However this approach
requires a substantial amount of data and its success depends on the
timescale over which the jitter is correlated, with shorter
correlation timescales being advantageous.

As we are mainly interested in the systems obliquity we only note
here that the relation by \cite{saar2003} employing a correlation
between $v \sin i_{\star}$ and stellar jitter, leads to an expected
jitter of $34$~m\,s$^{-1}$, similar to the measured value.

\section{Summary of conclusions}
\label{sec:conclusion}

We find that in the WASP-7 system the stellar spin axis is strongly
misaligned with the planet's orbital axis, by $86\pm6^\circ$ as
projected on the sky. This observation strengthens the correlation
found by \cite{winn2010} and lends support to the idea that systems
with close giant planets generally started out with a very broad range
of obliquities, and that the observed low obliquities of many systems
are a consequence of tidal dissipation.

Differential rotation and its imprint on the RM effect holds the
promise of measuring not only the projections of stellar obliquities,
but also the stellar inclinations. However, with the current
measurement precision and uncertainties in other parameters such as
limb darkening, no secure detection of differential rotation in WASP-7
can be made. We originally thought that WASP-7 might present a good
testbed to search for differential rotation via the RM effect, as the
misalignment of $90^{\circ}$ maximises the RM signal originating from
differential rotation. However for this angle the signal from
differential rotation is also strongly correlated with stellar limb
darkening. In addition WASP-7 displays a high degree of stellar
jitter. Therefore a system with a quiet star and a more moderate
misalignment might be better suited to search for signs of
differential rotation in the RM signal.

\acknowledgments S.A.\ thanks the Harvard-Smithsonian Center for
Astrophysics, where this manuscript was begun, for its
hospitality. S.A.\ acknowledges support by a Rubicon fellowship from
the Netherlands Organization for Scientific Research (NWO) during
parts of this project.  J.N.W.\ acknowledges support from a NASA
Origins grant (NNX09AD36G). T.H.\ is supported by Japan Society for
Promotion of Science (JSPS) Fellowship for Research (DC1:
22-5935). This research has made use of the Simbad database located at
{\tt http://simbad.u-strasbg.fr/}.


\begin{thebibliography}{38}
\expandafter\ifx\csname natexlab\endcsname\relax\def\natexlab#1{#1}\fi

\bibitem[{{Albrecht} {et~al.}(2007){Albrecht}, {Reffert}, {Snellen},
  {Quirrenbach}, \& {Mitchell}}]{albrecht2007}
{Albrecht}, S., {Reffert}, S., {Snellen}, I., {Quirrenbach}, A., \& {Mitchell},
  D.~S. 2007, \aap, 474, 565,
  \adsurl{http://adsabs.harvard.edu/abs/2007A\%26A...474..565A},
  \eprint{0708.2918}

\bibitem[{{Albrecht} {et~al.}(2011){Albrecht}, {Winn}, {Johnson}, {Butler},
  {Crane}, {Shectman}, {Thompson}, {Narita}, {Sato}, {Hirano}, {Enya}, \&
  {Fischer}}]{albrecht2011b}
{Albrecht}, S. {et~al.} 2011, \apj, 738, 50,
  \adsurl{http://adsabs.harvard.edu/abs/2011ApJ...738...50A},
  \eprint{1106.2548}

\bibitem[{{Butler} {et~al.}(1996){Butler}, {Marcy}, {Williams}, {McCarthy},
  {Dosanjh}, \& {Vogt}}]{butler1996}
{Butler}, R.~P., {Marcy}, G.~W., {Williams}, E., {McCarthy}, C., {Dosanjh}, P.,
  \& {Vogt}, S.~S. 1996, \pasp, 108, 500,
  \adsurl{http://adsabs.harvard.edu/abs/1996PASP..108..500B}

\bibitem[{{Chatterjee} {et~al.}(2008){Chatterjee}, {Ford}, {Matsumura}, \&
  {Rasio}}]{chatterjee2008}
{Chatterjee}, S., {Ford}, E.~B., {Matsumura}, S., \& {Rasio}, F.~A. 2008, \apj,
  686, 580, \adsurl{http://adsabs.harvard.edu/abs/2008ApJ...686..580C},
  \eprint{arXiv:astro-ph/0703166}

\bibitem[{{Claret}(2004)}]{claret2004}
{Claret}, A. 2004, \aap, 428, 1001,
  \adsurl{http://adsabs.harvard.edu/abs/2004A\%26A...428.1001C}

\bibitem[{{Collier Cameron} {et~al.}(2010){Collier Cameron}, {Bruce}, {Miller},
  {Triaud}, \& {Queloz}}]{cameron2010}
{Collier Cameron}, A., {Bruce}, V.~A., {Miller}, G.~R.~M., {Triaud},
  A.~H.~M.~J., \& {Queloz}, D. 2010, \mnras, 403, 151,
  \adsurl{http://adsabs.harvard.edu/abs/2010MNRAS.403..151C},
  \eprint{0911.5361}

\bibitem[{{Crane} {et~al.}(2010){Crane}, {Shectman}, {Butler}, {Thompson},
  {Birk}, {Jones}, \& {Burley}}]{crane2010}
{Crane}, J.~D., {Shectman}, S.~A., {Butler}, R.~P., {Thompson}, I.~B., {Birk},
  C., {Jones}, P., \& {Burley}, G.~S. 2010, in SPIE Conference Series, Vol.
  7735, 170, \adsurl{http://adsabs.harvard.edu/abs/2010SPIE.7735E.170C}

\bibitem[{{Cresswell} {et~al.}(2007){Cresswell}, {Dirksen}, {Kley}, \&
  {Nelson}}]{cresswell2007}
{Cresswell}, P., {Dirksen}, G., {Kley}, W., \& {Nelson}, R.~P. 2007, \aap, 473,
  329, \adsurl{http://adsabs.harvard.edu/abs/2007A\%26A...473..329C},
  \eprint{0707.2225}

\bibitem[{{Fabrycky} \& {Tremaine}(2007)}]{fabrycky2007}
{Fabrycky}, D., \& {Tremaine}, S. 2007, \apj, 669, 1298,
  \adsurl{http://adsabs.harvard.edu/abs/2007ApJ...669.1298F},
  \eprint{0705.4285}

\bibitem[{{Gaudi} \& {Winn}(2007)}]{gaudi2007}
{Gaudi}, B.~S., \& {Winn}, J.~N. 2007, \apj, 655, 550,
  \adsurl{http://adsabs.harvard.edu/abs/2007ApJ...655..550G}, \eprint{0608071}

\bibitem[{{Gray}(1984)}]{gray1984}
{Gray}, D.~F. 1984, \apj, 281, 719,
  \adsurl{http://adsabs.harvard.edu/abs/1984ApJ...281..719G}

\bibitem[{{Gray}(2005)}]{gray2005}
---. 2005, {The Observation and Analysis of Stellar Photospheres, 3$^{\rm rd}$
  Ed.} (ISBN 0521851866, Cambridge University Press),
  \adsurl{http://ukads.nottingham.ac.uk/cgi-bin/nph-bib_query?bibcode=2005oasp.book.....G&db_key=AST}

\bibitem[{{Hartman} {et~al.}(2011){Hartman}, {Bakos}, {Torres}, {Latham},
  {Kov{\'a}cs}, {B{\'e}ky}, {Quinn}, {Mazeh}, {Shporer}, {Marcy}, {Howard},
  {Fischer}, {Johnson}, {Esquerdo}, {Noyes}, {Sasselov}, {Stefanik},
  {Fernandez}, {Szklen{\'a}r}, {L{\'a}z{\'a}r}, {Papp}, \&
  {S{\'a}ri}}]{hartman2011}
{Hartman}, J.~D. {et~al.} 2011, ArXiv e-prints,
  \adsurl{http://adsabs.harvard.edu/abs/2011arXiv1106.1212H},
  \eprint{1106.1212}

\bibitem[{{Hatzes} {et~al.}(2010){Hatzes}, {Dvorak}, {Wuchterl}, {Guterman},
  {Hartmann}, {Fridlund}, {Gandolfi}, {Guenther}, \&
  {P{\"a}tzold}}]{hatzes2010}
{Hatzes}, A.~P. {et~al.} 2010, \aap, 520, A93,
  \adsurl{http://adsabs.harvard.edu/abs/2010A\%26A...520A..93H},
  \eprint{1006.5476}

\bibitem[{{H{\'e}brard} {et~al.}(2008){H{\'e}brard}, {Bouchy}, {Pont},
  {Loeillet}, {Rabus}, {Bonfils}, {Moutou}, {Boisse}, {Delfosse}, {Desort},
  {Eggenberger}, {Ehrenreich}, {Forveille}, {Lagrange}, {Lovis}, {Mayor},
  {Pepe}, {Perrier}, {Queloz}, {Santos}, {S{\'e}gransan}, {Udry}, \&
  {Vidal-Madjar}}]{hebrard2008}
{H{\'e}brard}, G. {et~al.} 2008, \aap, 488, 763,
  \adsurl{http://adsabs.harvard.edu/abs/2008A\%26A...488..763H},
  \eprint{0806.0719}

\bibitem[{{Hellier} {et~al.}(2009){Hellier}, {Anderson}, {Gillon}, {Lister},
  {Maxted}, {Queloz}, {Smalley}, {Triaud}, {West}, {Wilson}, {Alsubai},
  {Bentley}, {Cameron}, {Hebb}, {Horne}, {Irwin}, {Kane}, {Mayor}, {Pepe},
  {Pollacco}, {Skillen}, {Udry}, {Wheatley}, {Christian}, {Enoch}, {Haswell},
  {Joshi}, {Norton}, {Parley}, {Ryans}, {Street}, \& {Todd}}]{hellier2009}
{Hellier}, C. {et~al.} 2009, \apjl, 690, L89,
  \adsurl{http://adsabs.harvard.edu/abs/2009ApJ...690L..89H},
  \eprint{0805.2600}

\bibitem[{{Hirano} {et~al.}(2011{\natexlab{a}}){Hirano}, {Narita}, {Sato},
  {Winn}, {Aoki}, {Tamura}, {Taruya}, \& {Suto}}]{hirano2011b}
{Hirano}, T., {Narita}, N., {Sato}, B., {Winn}, J.~N., {Aoki}, W., {Tamura},
  M., {Taruya}, A., \& {Suto}, Y. 2011{\natexlab{a}}, ArXiv e-prints,
  \adsurl{http://adsabs.harvard.edu/abs/2011arXiv1108.4493H},
  \eprint{1108.4493}

\bibitem[{{Hirano} {et~al.}(2010){Hirano}, {Suto}, {Taruya}, {Narita}, {Sato},
  {Johnson}, \& {Winn}}]{hirano2010}
{Hirano}, T., {Suto}, Y., {Taruya}, A., {Narita}, N., {Sato}, B., {Johnson},
  J.~A., \& {Winn}, J.~N. 2010, \apj, 709, 458,
  \adsurl{http://adsabs.harvard.edu/abs/2010ApJ...709..458H},
  \eprint{0910.2365}

\bibitem[{{Hirano} {et~al.}(2011{\natexlab{b}}){Hirano}, {Suto}, {Winn},
  {Taruya}, {Narita}, {Albrecht}, \& {Sato}}]{hirano2011a}
{Hirano}, T., {Suto}, Y., {Winn}, J.~N., {Taruya}, A., {Narita}, N.,
  {Albrecht}, S., \& {Sato}, B. 2011{\natexlab{b}}, ArXiv e-prints,
  \adsurl{http://adsabs.harvard.edu/abs/2011arXiv1108.4430H},
  \eprint{1108.4430}

\bibitem[{{Johnson} {et~al.}(2009){Johnson}, {Winn}, {Albrecht}, {Howard},
  {Marcy}, \& {Gazak}}]{johnson2009}
{Johnson}, J.~A., {Winn}, J.~N., {Albrecht}, S., {Howard}, A.~W., {Marcy},
  G.~W., \& {Gazak}, J.~Z. 2009, \pasp, 121, 1104,
  \adsurl{http://adsabs.harvard.edu/abs/2009PASP..121.1104J},
  \eprint{0907.5204}

\bibitem[{{Laughlin} {et~al.}(2005){Laughlin}, {Marcy}, {Vogt}, {Fischer}, \&
  {Butler}}]{laughlin2005}
{Laughlin}, G., {Marcy}, G.~W., {Vogt}, S.~S., {Fischer}, D.~A., \& {Butler},
  R.~P. 2005, \apjl, 629, L121,
  \adsurl{http://adsabs.harvard.edu/abs/2005ApJ...629L.121L}

\bibitem[{{Lin} {et~al.}(1996){Lin}, {Bodenheimer}, \& {Richardson}}]{lin1996}
{Lin}, D.~N.~C., {Bodenheimer}, P., \& {Richardson}, D.~C. 1996, \nat, 380,
  606, \adsurl{http://adsabs.harvard.edu/abs/1996Natur.380..606L}

\bibitem[{{Mandel} \& {Agol}(2002)}]{mandel2002}
{Mandel}, K., \& {Agol}, E. 2002, \apjl, 580, L171,
  \adsurl{http://adsabs.harvard.edu/abs/2002ApJ...580L.171M}, \eprint{0210099}

\bibitem[{{Maxted} {et~al.}(2011){Maxted}, {Koen}, \& {Smalley}}]{maxted2011}
{Maxted}, P.~F.~L., {Koen}, C., \& {Smalley}, B. 2011, ArXiv e-prints,
  \adsurl{http://adsabs.harvard.edu/abs/2011arXiv1108.0349M},
  \eprint{1108.0349}

\bibitem[{{Nagasawa} {et~al.}(2008){Nagasawa}, {Ida}, \&
  {Bessho}}]{nagasawa2008}
{Nagasawa}, M., {Ida}, S., \& {Bessho}, T. 2008, \apj, 678, 498,
  \adsurl{http://adsabs.harvard.edu/abs/2008ApJ...678..498N},
  \eprint{0801.1368}

\bibitem[{{Ohta} {et~al.}(2005){Ohta}, {Taruya}, \& {Suto}}]{ohta2005}
{Ohta}, Y., {Taruya}, A., \& {Suto}, Y. 2005, \apj, 622, 1118,
  \adsurl{http://adsabs.harvard.edu/cgi-bin/nph-bib_query?bibcode=2005ApJ...622.1118O&db_key=AST},
  \eprint{astro-ph/0410499}

\bibitem[{{Pont} {et~al.}(2011){Pont}, {Husnoo}, {Mazeh}, \&
  {Fabrycky}}]{pont2011}
{Pont}, F., {Husnoo}, N., {Mazeh}, T., \& {Fabrycky}, D. 2011, \mnras, 414,
  1278, \adsurl{http://adsabs.harvard.edu/abs/2011MNRAS.414.1278P},
  \eprint{1103.2081}

\bibitem[{{Reiners} \& {Schmitt}(2003)}]{reiners2003}
{Reiners}, A., \& {Schmitt}, J.~H.~M.~M. 2003, \aap, 398, 647,
  \adsurl{http://adsabs.harvard.edu/cgi-bin/nph-bib_query?bibcode=2003A\%26A...398..647R&db_key=AST}

\bibitem[{{Saar} {et~al.}(2003){Saar}, {Hatzes}, {Cochran}, \&
  {Paulson}}]{saar2003}
{Saar}, S.~H., {Hatzes}, A., {Cochran}, W., \& {Paulson}, D. 2003, in The
  Future of Cool-Star Astrophysics: 12th Cambridge Workshop on Cool Stars,
  Stellar Systems, and the Sun, ed. {A.~Brown, G.~M.~Harper, \& T.~R.~Ayres},
  Vol.~12, 694--698, \adsurl{http://adsabs.harvard.edu/abs/2003csss...12..694S}

\bibitem[{{Schlaufman}(2010)}]{schlaufman2010}
{Schlaufman}, K.~C. 2010, \apj, 719, 602,
  \adsurl{http://adsabs.harvard.edu/abs/2010ApJ...719..602S},
  \eprint{1006.2851}

\bibitem[{{Shporer} \& {Brown}(2011)}]{shporer2011}
{Shporer}, A., \& {Brown}, T. 2011, \apj, 733, 30,
  \adsurl{http://adsabs.harvard.edu/abs/2011ApJ...733...30S},
  \eprint{1103.0775}

\bibitem[{{Simpson} {et~al.}(2011){Simpson}, {Pollacco}, {Cameron},
  {H{\'e}brard}, {Anderson}, {Barros}, {Boisse}, {Bouchy}, {Faedi}, {Gillon},
  {Hebb}, {Keenan}, {Miller}, {Moutou}, {Queloz}, {Skillen}, {Sorensen},
  {Stempels}, {Triaud}, {Watson}, \& {Wilson}}]{simpson2011}
{Simpson}, E.~K. {et~al.} 2011, \mnras, 414, 3023,
  \adsurl{http://adsabs.harvard.edu/abs/2011MNRAS.414.3023S},
  \eprint{1011.5664}

\bibitem[{{Southworth} {et~al.}(2011){Southworth}, {Dominik}, {J{\o}rgensen},
  {Rahvar}, {Snodgrass}, {Alsubai}, {Bozza}, {Browne}, {Burgdorf}, {Calchi
  Novati}, {Dodds}, {Dreizler}, {Finet}, {Gerner}, {Hardis}, {Harps{\o}e},
  {Hellier}, {Hinse}, {Hundertmark}, {Kains}, {Kerins}, {Liebig}, {Mancini},
  {Mathiasen}, {Penny}, {Proft}, {Ricci}, {Sahu}, {Scarpetta}, {Sch{\"a}fer},
  {Sch{\"o}nebeck}, \& {Surdej}}]{southworth2011}
{Southworth}, J. {et~al.} 2011, \aap, 527, A8,
  \adsurl{http://adsabs.harvard.edu/abs/2011A\%26A...527A...8S},
  \eprint{1012.5181}

\bibitem[{{Triaud}(2011)}]{triaud2011}
{Triaud}, A.~H.~M.~J. 2011, \aap, 534, L6,
  \adsurl{http://adsabs.harvard.edu/abs/2011A\%26A...534L...6T},
  \eprint{1109.5813}

\bibitem[{{Triaud} {et~al.}(2010){Triaud}, {Collier Cameron}, {Queloz},
  {Anderson}, {Gillon}, {Hebb}, {Hellier}, {Loeillet}, {Maxted}, {Mayor},
  {Pepe}, {Pollacco}, {S{\'e}gransan}, {Smalley}, {Udry}, {West}, \&
  {Wheatley}}]{triaud2010}
{Triaud}, A.~H.~M.~J. {et~al.} 2010, \aap, 524, A25,
  \adsurl{http://adsabs.harvard.edu/abs/2010A\%26A...524A..25T},
  \eprint{1008.2353}

\bibitem[{{Winn} {et~al.}(2010){Winn}, {Fabrycky}, {Albrecht}, \&
  {Johnson}}]{winn2010}
{Winn}, J.~N., {Fabrycky}, D., {Albrecht}, S., \& {Johnson}, J.~A. 2010, \apjl,
  718, L145, \adsurl{http://adsabs.harvard.edu/abs/2010arXiv1006.4161W},
  \eprint{1006.4161}

\bibitem[{{Winn} {et~al.}(2011){Winn}, {Howard}, {Johnson}, {Marcy},
  {Isaacson}, {Shporer}, {Bakos}, {Hartman}, {Holman}, {Albrecht}, {Crepp}, \&
  {Morton}}]{winn2010b}
{Winn}, J.~N. {et~al.} 2011, \aj, 141, 63,
  \adsurl{http://adsabs.harvard.edu/abs/2011AJ....141...63W},
  \eprint{1010.1318}

\bibitem[{{Winn} {et~al.}(2009){Winn}, {Johnson}, {Albrecht}, {Howard},
  {Marcy}, {Crossfield}, \& {Holman}}]{winn2009}
{Winn}, J.~N., {Johnson}, J.~A., {Albrecht}, S., {Howard}, A.~W., {Marcy},
  G.~W., {Crossfield}, I.~J., \& {Holman}, M.~J. 2009, \apjl, 703, L99,
  \adsurl{http://adsabs.harvard.edu/abs/2009ApJ...703L..99W},
  \eprint{0908.1672}

\end{thebibliography}

\end{document}